\documentclass[]{interact}

\usepackage{epstopdf}% To incorporate .eps illustrations using PDFLaTeX, etc.
\usepackage{subfigure}% Support for small, `sub' figures and tables
 
\usepackage{amsmath,bm,amssymb}
\usepackage{graphicx,xcolor,soul}
\usepackage{tabularx,ragged2e,booktabs}

\usepackage[colorlinks = true,linkcolor = blue,urlcolor  = cyan,citecolor = red,anchorcolor = blue]{hyperref}

\usepackage[numbers,sort&compress,merge]{natbib}
\usepackage{notoccite}

\newcommand{\px}{\hat{\sigma}_x}

\newcommand{\pz}{\hat{\sigma}_z}

\newcommand{\bra}[1]{\langle #1 |}
\newcommand{\ket}[1]{| #1\rangle}
\begin{document}

%\articletype{Draft:- Results}
\articletype{REGULAR ARTICLE}

\title{Towards Laser Control of Open Quantum Systems: Memory Effects}

\author{
\name{R.\;Puthumpally-Joseph\textsuperscript{1,2}, O.\;Atabek\textsuperscript{2},
E.\;Mangaud\textsuperscript{3},
M.\;Desouter-Lecomte\textsuperscript{4,5}\thanks{Contact M.\;Desouter-Lecomte. E-mail: michele.desouter-lecomte@u-psud.fr}, 
D.\;Sugny\textsuperscript{1,6}
}
\affil{\textsuperscript{1}Laboratoire Interdisciplinaire Carnot de Bourgogne (ICB), UMR 6303 CNRS-Universit\'e Bourgogne Franche Comt\'e, 9 Av. A. Savary, BP 47 870, F-21078 Dijon cedex, France
\newline
\textsuperscript{2}Institut des Sciences Mol\'eculaires d'Orsay (ISMO), Univ. Paris-Sud, CNRS, Universit\'e Paris-Saclay, 91405 Orsay cedex, France
\newline
\textsuperscript{3}Institut f\"ur Physikalische und Theoretische Chemie, Julius-Maximilians Universit\"at W\"urzburg, Emil-Fischer-Strasse 42, 97074, W\"urzburg, Germany
\newline
\textsuperscript{4}Laboratorie Chimie Physique (LCP) , Univ. Paris-Sud, CNRS, Universit\'e Paris-Saclay, 91405 Orsay cedex, France
\newline
\textsuperscript{5}D\'epartement de Chimie, Universit\'e de Li\`ege, Sart Tilman, B6, B-4000 Li\`ege, Belgium
\newline
\textsuperscript{6}Institute for Advanced Study, Technische Universit\"at M\"unchen, Lichtenbergstrasse 2 a, D-85748 Garching, Germany
\newline
}}

\maketitle

\begin{pacscode} 33.80.-b, 03.65.Yz, 42.50.Hz  \end{pacscode}
\begin{abstract}
Laser control of Open Quantum Systems (OQS) is a challenging issue as compared to its counterpart in isolated small size molecules, basically due to very large numbers of degrees of freedom to be accounted for. Such a control aims at appropriately optimizing decoherence processes of a central two-level system (a given vibrational mode, for instance) towards its environmental bath (including, for instance, all other normal modes). A variety of applications could potentially be envisioned, either to preserve the central system from decaying (long duration molecular alignment or orientation, qubit decoherence protection) or, to speed up the information flow towards the bath (efficient charge or proton transfers in long chain organic compounds). Achieving such controls require some quantitative measures of decoherence in relation with memory effects in the bath response, actually given by the degree of non-Markovianity. Characteristic decoherence rates of a Spin-Boson model are calculated using a Nakajima-Zwanzig type master equation with converged HEOM expansion for the memory kernel. It is shown that, by adequately tuning the two-level transition frequency through a controlled Stark shift produced by an external laser field, non-Markovianity can be enhanced in a continuous way leading to a first attempt towards the control of OQS.  
\end{abstract}

%!!!!!!!!!!!!!!!!!!!!!!!!!!!!!!!!!!!!!!!!!!!!!!!!!!!!!! %
\section{Introduction}

The theory of Open Quantum Systems (OQS) deals with non-unitary dynamics among quantum states of a sub-system coupled with some unobservable degrees of freedom in their surroundings \cite{breuerbook, weissbook, RMPVega}. This is actually a very common situation since no physical system can truly be considered as isolated. The additional degrees of freedom that make up the environment can take many physical forms such as phonons in solids, photons in cavities, charge fluctuations, or molecular vibrations. They are responsible for a wide variety of basic processes and, in particular, thermalization, energy and charge transfers, decoherence. As a consequence, the scope of OQS encompasses several disciplines of both fundamental and technological importance, from struggling against decoherence in superconducting qubits of quantum information, to electronic and proton transfers in flexible proteins. This is why OQS theory has attracted considerable attention both from physics and chemistry communities.

Understanding non-equilibrium dynamics in OQS where dissipation and coherence evolve simultaneously is by itself a great challenge, but even more important is the depiction of feasible quantum control strategies to optimize physical observables such as decoherence rates or efficient and fast charge transfers over large molecules. As an illustrative example, non-linear ultra-fast optics experiments have shown the possibility to protect from environmental fluctuations, on picosecond time scales, photosynthetic organisms in molecular light harvesting materials \cite{Chin2013, Gelinas2014}. This has even led to the emergence of a new interdisciplinary field: Quantum Biology \cite{Scholes2011}. In this work we are concerned by a strong laser control scheme addressing OQS described in a standard topology of a central system of few degrees of freedom coupled to an external and much larger set of environmental modes, namely, the residual bath which damps the dynamics. 
This problem could naively be approached by a time-dependent wave function evolution incorporating both the central system and the bath. But its feasibility would be very soon affected with the increasing number of degrees of freedom. It is important to have in mind that even referring to high performance variational codes like MCTDH with multi-layer, time-adapted basis set expansions, but only a few hundreds of such vibrational modes can reasonably be taken into account \cite{GattiMCTDH}. Stochastic methods would also be limited to comparable numbers of degrees of freedom \cite{stochastic}. A strong field Floquet type of approach, as has been applied to H$_2^+$ photodissociation by Andr\'e Bandrauk and coworkers, would even be worse, as the resulting number of close coupled equations to describe the multiphoton excitation process would lead to computationally non tractable issues \cite{Bandrauk}. One of the goals of OQS theory is precisely to avoid a full integration of all the degrees of freedom by describing the dynamics in a reduced Hilbert space, through a reduced density matrix. More specifically, such a reduction is obtained by tracing out the bath degrees of freedom from the full density matrix.  
In some cases the damping could be an almost pure, memory-less dissipation.The dynamics is then approximately described referring to Markovian master equations in the so-called Lindblad form \cite{Lindblad1, Lindblad2} extensively used in atomic physics, quantum optics, semiconductors. This approximation assumes an instantaneous recovery of the bath from the interaction with a continuous flow of information from the central system to its environment. But, when aiming at control OQS at different times and length scales, or different energy and temperature ranges, dissipation and decoherence dynamics could be such that the bath response to the central system can no longer be neglected, leading thus to non-Markovian behaviours characterized by a back-flow of information. 

Two classes of control strategies can be devised: either acting on the central system alone, or on both the central system and its environmental bath. The present work is concerned by the simplest scenario where a strong static field produces a Stark shift affecting the eigen-energies of a two-level central system leading to different coupling schemes with the bath. A more complete scenario would be the one which takes advantage of the non-Markovian response of the bath to control the full system dynamics \cite{KochReview}. Such an ultimate goal could presumably be reached by referring to some collective modes which guide the flow of information from the central system to the bath in a reversible manner \cite{collectivemode}. But we still expect that the Stark shift affecting the central system can be used as a first step for enhancing non-Markovian behaviours. Obviously a requisite of any control strategy exploiting non-Markovianity is a quantitative measure of such a behaviour. Several measures characterizing non-Markovianity have been proposed in the literature \cite{tracedistance, tracedistance2, PhysRevA.88.020102, relaxrates1}, among them the volume of accessible states \cite{PhysRevA.88.020102} and more recently the appearance of negative decoherence rates from several possible decoherence channels \cite{relaxrates1}. The relative merits of such measures based on the quantum dynamical map (with a matrix representation $F$) of the reduced density matrix elements deserve interest. More specifically, the volume of accessible states within the Bloch sphere, given by the determinant of $F$, evolves in time with a total rate $\Gamma(t)$. The time-dependence of $\Gamma(t)$, as opposite to a constant value, is a signature of memory effects. But an even more sensitive signature would be reached through the partial decoherence rates $\gamma_k(t)$ towards the different decoherence channels $k$ of the dynamics. Their calculation is based on, roughly speaking, a logarithmic derivative of $F$, diagonalizing $\dot{F}F^{-1}$. These partial rates sum up to $\Gamma(t)$ but could individually and temporarily be negative, as signatures of information back-flow from the bath to the central system. We are postponing their detailed analysis to a forthcoming paper.

The article is organized as follows.  
Section \ref{Sec:theo} is devoted to the theory describing the dynamics of a two-level system coupled to a bosonic bath, the so-called Spin-Boson model. A canonical (system Hamiltonian-independent) form for a time-local master equation is introduced, for a \emph{generic} $F(t)$ mapping matrix. This is done by closely following the derivation of Ref.\cite{relaxrates1}. The system-specific reduced density matrix evolution and the resulting \emph{physical} $F(t)$ is calculated by numerically solving a Nakajima-Zwanzig type of evolution using hierarchical equations of motion (HEOM) up to convergence, for the memory kernel \cite{heom1, heom2, jcp126.2538754, jcp130.3155372, PhysRevE.86.011915}. The results are discussed in Section \ref{Sec:Results} in terms of the time-evolution of an observable taken as the volume of states decaying with $\Gamma(t)$. The most important observation is the enhancement of the non-Markovian behaviour when the two-level system internal transition frequency is off-resonance with respect to the maximum of the spectral density measuring the frequency distribution of the system-bath couplings. The control knob is then the external field strength inducing the Stark shift that monitors the system transition frequency (from on- to off-resonance). At this stage, the control targets the optimization of decoherence rates in order to slow down the overall decay, which by itself opens important applications in protecting quantum information lost or obtaining long-lived alignment-orientation in molecular dynamics (although not described by a Spin-Boson Hamiltonian), for instance. Future perspectives on other control goals aiming at adequately designing collective modes in the bath could open the way for efficient charge or proton transfer processes in large molecular systems. These are mentioned in the conclusion, Section \ref{Sec:Conclusion}.

\section{Theory and Method}
\label{Sec:theo}
%!!!!!!!!!!!!!!!!!!!!!!!!!!!!!!!!!!!!!!!!!!!!!!!!!!!!!! %

\subsection{Canonical Master Equation}
In many situations, in particular
in liquid and solid state physics, memory effects due to the
non-Markovian character of the dynamics have to be taken into account \cite{RMPbreuer,reviewrivas}. At this point, a natural
question about a given experimental system is to which extent the relaxation effects lead to non-Markovian dynamics. In the past few years, the problem of how to measure non-Markovianity has sparked remarkable interest, starting with Refs. \cite{PhysRevLett.105.050403,tracedistance,tracedistance2,PhysRevLett.105.050403}. Number of quantitative measures have been proposed since these initial attempts \cite{PhysRevA.88.020102,relaxrates1,RMPbreuer,PhysRevA.84.052118,PhysRevA.87.042103,PhysRevA.82.042103,PhysRevA.82.042103,PhysRevLett.112.120404}. In addition to the theoretical studies, recent experimental works \cite{mesure_expnature,mesure_expnaturephys} have shown that it is by now possible to engineer OQS and to drive it from the Markovian to the non-Markovian regimes. However, the measures recently proposed in the literature are well-suited to theoretical or abstract situations,
where for example the dynamics can be solved analytically. Very few papers have explored more complicated model systems where the time evolution of the OQS can only be achieved numerically \cite{tracedistanceSB,nonmarkoviansaalfrank, JChemPhys.134.101103}. A benchmark example in this category is the spin boson model which will be analyzed in this paper. We will consider for this model system a specific measure of non-Markovianity, namely the volume of accessible state space \cite{relaxrates1,relaxrates2}. The goal of this paragraph is to briefly summarize the computation of this measure. The reader is referred to the original papers for technical details.

The dynamics of a Markovian open quantum system is typically ruled by a quantum dynamical semi-group \cite{breuerbook}. The most general representation of such semigroup is given by the Lindblad-Kossakowski equation \cite{Lindblad1,Lindblad2} which can be expressed as:
\begin{equation}
\frac{d}{dt}\rho_S(t)=\mathcal{L}\rho_S(t),
\end{equation}
where $\rho_S(t)$ is the density matrix of the central system under study and $\mathcal{L}$ a generator in Lindblad form. This latter can be written as:
\begin{equation}
\mathcal{L}\rho_S(t)=-i[H(t),\rho_S(t)]+\sum_i \gamma_i [A_i\rho_S A_i^\dagger-\frac{1}{2}\{A_i^\dagger A_i,\rho_S\}],
\end{equation}
$H$ being the Hamiltonian of the system, the coefficients $\gamma_i\geq 0$ the relaxation rates and the $A_i$ the time-independent Lindblad operators. Hereafter, we assume that $\rho_S(t)$ is a density matrix of a finite-dimensional Hilbert space, typically of dimension two for a spin system. In this case, the dynamics is characterized by three relaxation rates, $\gamma_i$.

This formalism can be generalized to a time-local master equation where the generator $\mathcal{L}$ depends on time while preserving its general form:
\begin{eqnarray}
& &\mathcal{L}\rho_S(t)=-i[H(t),\rho_S(t)]\\
& & +\sum_i \gamma_i(t) [A_i(t)\rho_S A_i^\dagger (t)-\frac{1}{2}\{A_i^\dagger (t)A_i(t),\rho_S\}].\nonumber
\end{eqnarray}
We stress here that both the relaxation rates $\gamma_i$ and the Lindblad operators $A_i(t)$ may now depend on time. The process remains Markovian if the rates are positive for any time $t$, and presumably becomes non-Markovian otherwise \cite{relaxrates1}.

Recent studies have investigated the issue of the description of the dynamics of non-Markovian systems which can be given either by a non-local master equation with memory kernel obtained, e.g., from the Nakajima-Zwanzig technique \cite{breuerbook}, or by a local in time equation \cite{relaxrates1, localnonlocal}. We give here a simple argument showing that any time evolution of the density matrix of the system can be ruled by a time-local equation. We introduce the quantum dynamical map $F(t)$ of the system which satisfies:
\begin{equation}\label{eqf}
\rho_S(t)=F(t)\rho_S(0),~t\geq 0.
\end{equation}
Differentiating Eq.~(\ref{eqf}) with respect to time leads to:
\begin{equation}\label{eqlind}
\dot{\rho_S}(t)=\dot{F}(t)\rho_S(0).
\end{equation}
If we assume that the dynamical map is not singular at time $t$, i.e. that the determinant of the corresponding matrix is different from 0, then the inverse $F^{-1}$ of $F$ can be defined and Eq.~(\ref{eqlind}) can be rewritten as follows:
\begin{equation}
\dot{\rho_S}=\mathcal{L}(t)\rho_S(t)=\dot{F}F^{-1}\rho_S(t),
\end{equation}
which is a master equation local in time. The non-Markovian character of the dynamics is then associated with the values of the relaxation rates of the generator $\mathcal{L}$.

$F(t)$ is also referred to when mapping the density matrix on its corresponding Bloch ball using Pauli matrices together with the identity as an orthogonal basis set.  
The time evolution of the volume of this Bloch ball $V(t)$, namely the volume of accessible states, is given by the determinant of $F(t)$:
\begin{equation}
	V(t) = det [ F(t) ]  \,,
	\label{Eq:Vol_Access_state}
\end{equation}	
which constitutes another signature of non-Markovian dynamics \cite{PhysRevA.88.020102}. 
More precisely, the dynamics is said to be non-Markovian if:
\begin{equation}
\frac{d V(t)}{dt}>0
\end{equation}
for a given time $ t $.	
In this work we will merely focus on this signature, leaving the calculation of partial relaxation rates and their possible negativity for a forthcoming paper.

\subsection{Hierarchical equations of motion}
We consider a spin-boson Hamiltonian currently used to model a lot of processes such as electron or proton transfer, excitation energy transfer and qubit in a surrounding
\begin{subequations}
\label{Eq:hamiltonian}
\begin{eqnarray}
\hat{H} & = & \hat{H}_S + \hat{H}_B + \hat{H}_{SB} + \hat{H}_{\rm{ren}}\\
& = & \dfrac{{\omega}^d_0}{2\hbar} \pz + W \px + \dfrac{1}{2}\sum_{j}^N \hat{p}_{j}^{2} + \omega_{j}^{2} \left( \hat{q}_{j} - \dfrac{c_{j}}{\omega_{j}^{2} }\pz\right)^{2} \,,
\end{eqnarray}
\end{subequations}
where,
\begin{equation}
\hat{H}_S = \dfrac{{\omega}^d_0}{2\hbar} \pz + W \px 
\end{equation}
and
\begin{equation}
\hat{H}_B + \hat{H}_{SB} + \hat{H}_{\rm{ren}} = \dfrac{1}{2}\sum_{j} \hat{p}_{j}^{2} + \omega_{j}^{2} \left( \hat{q}_{j} - \dfrac{c_{j}}{\omega_{j}^{2} }\pz\right)^{2} \,.
\end{equation}
$ \{\hat{\sigma}_{\alpha}\} $ are Pauli matrices ($ \pz =\ket{1}_d\;{}_d\bra{1} -\ket{2}_d\;{}_d\bra{2} $ and $  \px = \ket{1}_d\;{}_d\bra{2} +\ket{2}_d\;{}_d\bra{1}$) written in a zeroth order, so called diabatic basis labelled $ d $. $W$ is the central system  interstate coupling in a diabatic representation and the vibrational modes are written in mass weighted coordinates. Adopting an adiabatic representation for the central system, by diagonalizing the system Hamiltonian $ \hat{H}_S $ results into the central system transition frequency $\omega_0= 2\sqrt{{\omega_0^d}^2/4 + W^2}$, which is hereafter taken as the single parameter defining the central system.  The system-bath coupling takes the form
\begin{equation}
\hat{H}_{SB} = -\sum_{j} c_{j}\hat{q}_{j}\pz \quad = -\hat{Q}\pz \quad = -\hat{Q}\hat{S}\,,
\label{Eq:HSBCouple}
\end{equation}
where $\hat{Q}= \sum_{j} c_{j}\hat{q}_{j} $ is a collective bath coordinate which couples to the system Hamiltonian by inducing fluctuations of the energy gap and $ \hat{S} $ is the generic central system coordinates. $ \hat{H}_{\rm{ren}} = \pz\,(c_{j}/\sqrt{2}\omega_{j})^{2} $ is the renormalization energy shifting the system energies.
The key tool in dissipative dynamics is the reduced density matrix which is the partial trace of the full density matrix over the bath degrees of freedom. 
\begin{equation}
\rho_{S}(t) = {\rm Tr}_{B} \left[\rho(t)\right]\,.
\label{Eq:Reduced_Density_Mat_S}
\end{equation}
Quantum information is exchanged between the system and the environment causing energy relaxation and decoherence.

%%%%%%%%%%%%%%% OA    %%%%%%%%%%%%%%%%%%

Well known projection techniques leading to non-Markovian master equations either time non local in the Nakajima-Zwanzig approach \cite{Nakajima01121958, ZWANZIG19641109}  or time local in the Hashitsumae-Shibata-Takahashi formalism \cite{Shibata1977,Hashitsumae1977} provide effective equations for the relevant central system part 
\begin{equation}
\partial_t\rho_S(t)=\mathcal{L}_{\rm eff}(t)\rho_S(t)+\int_{0}^{t}dt'\mathcal{K}(t,t')\rho_S(t')\,,
\label{Eq:MasterEqn}
\end{equation}
where
\begin{equation}
\mathcal{L}_{\rm eff}(t)\rho_S(t) = -i \left[ \hat{H}_S (t) + \hat{H}_{\rm ren}, \rho_S(t)\right]\,.
\end{equation}
At the initial time ($ t=0 $), the following factorization is assumed:
\begin{equation}
\rho(0) = \rho_S(0) \rho_{eq}\,,
\end{equation}
where $\rho_{eq}$ is the density matrix of the bath at equilibrium.

The exact reduced density matrix can in principle be obtained from the Feynman-Vernon influence functional \cite{FEYNMAN1963118} and has been implemented in some simple cases \cite{Makri95, Makri99}. However the evaluation of the  memory kernel $\mathcal{K}(t,t')$, in practical applications, remain cumbersome for complex systems. For more numerical efficiency different strategies belonging to two main classes, projection techniques or hierarchy equations of motions, have addressed differential equations of motion. HEOM originally proposed by Tanimura and Kubo \cite{heom1, PhysRevA.41.6676, heom2} is a powerful method providing a non-perturbative calculation of $ \rho_{S}(t) $ in Markovian or non-Markovian baths. Details about the derivation of the HEOM equations can be found in the original papers and in different reviews \cite{JChemPhys.122.041103, JChemPhys.122.041103}. It is born from the path integral method in the case of a harmonic bath allowing for analytical treatment of the influence functional, in particular the derivation of a hierarchy of time local coupled equations when the correlation function of the collective bath mode can be expanded as a sum of exponential functions. The correlation function of the collective coordinate is a Boltzmann average over the equilibrium bath ensemble defined by
\begin{equation}
\mathcal{C}(t) = {\rm Tr}_B \left[\hat{Q}(t)\hat{Q}(0)\rho_{B}\right]\,,
\label{Eq:Correlation1}
\end{equation}
where $\hat{Q}(t) = e^{i\hat{H}_B t/\hbar} \hat{Q} e^{-i\hat{H}_B t/\hbar}$ is the Heisenberg representation of the bath coordinate and $\rho_B$ the bath density matrix. The fluctuation-dissipation theorem relates the correlation function to a spectral density $\mathcal{J}(\omega)$ \cite{weissbook}
\begin{equation}
\mathcal{C}(t,t_0) =\frac{1}{\pi} \int_{-\infty}^{\infty} \dfrac{e^{-i\omega(t-t_0)}}{1-e^{-\beta\hbar\omega}} \,\mathcal{J}(\omega)d\omega\,,
\label{Eq:Corr_fn}
\end{equation}
 where $\beta=1/k_B T$ and $k_B$ the Boltzman factor.

Different schemes have been proposed to decompose the spectral density \cite{jcp140.4870035} which is nothing but a frequency distribution of the system-bath coupling. We discuss the operational equations in the case where the spectral density is expanded as a sum of Lorentzian functions with Ohmic behaviour at low frequencies \cite{tannor_meier_jw, Pomyalov201098}
\begin{equation}
\mathcal{J}(\omega) = \sum_{l=1}^{M} \dfrac{\omega \Delta_{l}}{\left[(\omega-\omega_l)^2 + \Omega_{l}^2\right]\left[(\omega+\omega_l)^2 + \Omega_{l}^2\right]}\,,
\label{Eq:jw}
\end{equation}
$\Delta_l$ being the coupling amplitude of the central system to the $l$th labelled collective-bath mode.
Referring to Cauchy residues theorem, the correlation function takes the form of an exponential series involving both the poles of $\mathcal{J}(\omega)$ and those of the Bose function through Matsubara frequencies \cite{NIETO199554} with complex $ \alpha_k $ and $ \zeta_k $
\begin{equation}
\mathcal{C}(t) = \sum_{k}^{K} \alpha_ke^{i\zeta_k t}\,.
\end{equation}
 The complex conjugate correlation function can be written by using the same $ \zeta_k $ but different expansion coefficients $ \tilde{\alpha}_k $  \cite{Pomyalov201098}
\begin{equation}
\mathcal{C}^*(t) = \sum_{k}^{K} \tilde{\alpha}_ke^{i\zeta_k t}\,,
\end{equation}
where $ K $ being the total number of dissipative modes
The reduced system density matrix is then the first element of a chain of auxiliary density matrices $ \rho_{\bm n} (t) $. The evolution is driven by time local coupled equations:
\begin{eqnarray}
\dot{\rho}_{\bm n} &=& -i\left[\hat{H}_{S}, \rho_{\bm n}(t)\right]\nonumber \\
&+& i\sum_k n_k \zeta_k\rho_{\bm n}(t)-i\left[\hat{S}, \sum_k \rho_{{\bm n}_k^+}(t)\right]-i\sum_k n_k \left(\alpha_k \hat{S} \rho_{{\bm n}_k^-}-\tilde{\alpha}_k  \rho_{{\bm n}_k^-}\hat{S}\right).
\label{Eq:HEOM_aux_mat}
\end{eqnarray}
The hierarchy is built with $ L $ levels. Each level corresponds to a given sum of the occupation numbers $n_k$,  $ L = \sum_{k}^{K} n_k $ of the $ K $ dissipation modes. Level $  L = 0 $ corresponds to the system matrix $ \rho(t) = \rho_{\bm 0}(t) $ with the vector $ \bm 0 = \{0,0,\cdots,0\} $, i.e. all $n_k$'s are zero. Each density matrix of level $ L $ is coupled to matrices of level $ L\pm 1 $. This very efficient algorithm only uses a single correlation function to describe the system bath interaction. Moreover, the structure is well suited for implementation on parallel computers \cite{JChemTeoComp.8.ct3003833}. 
In the following Eq.~(\ref{Eq:MasterEqn}) has been approximately solved at a given level of hierarchy $L$ corresponding to $ 2L $ perturbation order, by an appropriate truncation of  Eq.~(\ref{Eq:HEOM_aux_mat})

\subsection{The Model.}

\begin{figure}[ht]
	\centering
	\includegraphics{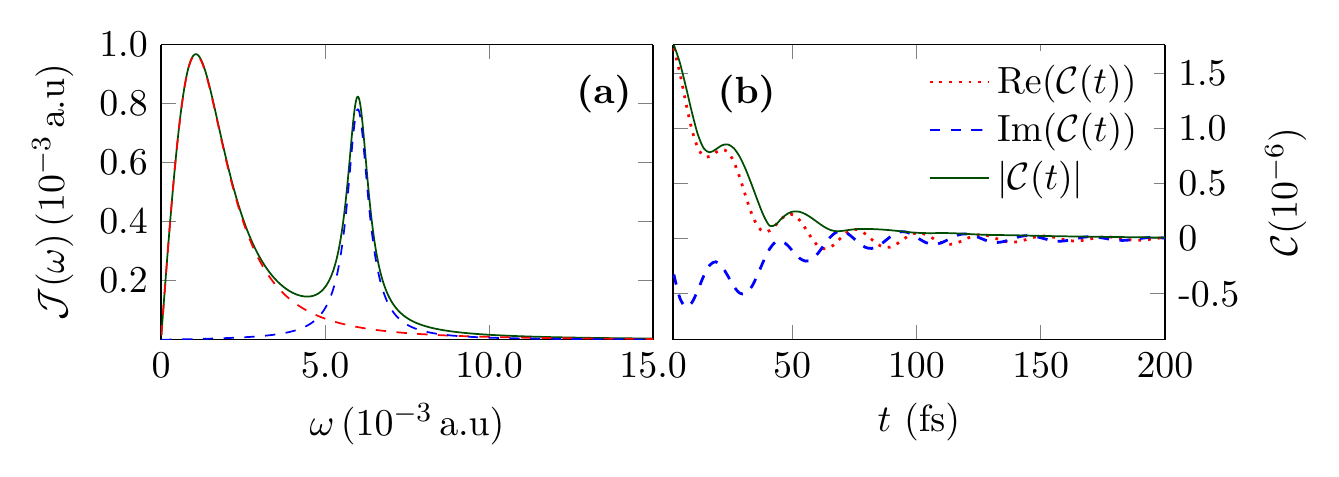}
	\caption{(Color online)  (a):\:Spectral density for the parameters given in table\:\ref{table:SD_values}. Red and blue dashed curves are the two antisymmetrised Lorentzian components of $ \mathcal{J}(\omega) $. (b):\:Correlation function associated with $ \mathcal{J}(\omega) $. }
	\label{Fig:jw_ct}
\end{figure}

As been discussed in the previous paragraph the central part of the Spin-Boson system is basically modeled by a single parameter two-level system, namely its internal transition frequency $\omega_0$ defining the energy gap $\hbar\omega_0$ between states $|1\rangle$ and $|2\rangle$. Note that in some cases, this could result from the diagonalization of a primary so-called diabatic two levels $|1\rangle_d$ and $|2\rangle_d$ directly interacting through a potential coupling $W$. The bosonic bath is described in terms of $N$ harmonic oscillators of frequency $\omega_j$ ($j=1,2,...N$). As for the coupling of the central system to the bath, it is taken into account referring to a spectral density, $\mathcal{J}(\omega)$. It is worthwhile noting that, in the absence of an external field, the individual levels making up the central system are only indirectly coupled through their environmental bath. By analogy with a standard Fano model of two discrete levels facing and interacting with a continuum, one can refer to $\mathcal{J}(\omega)$ as a frequency representation of an energy-dependent discrete-continuum coupling scheme appropriately averaged on the density of levels of the discretized continuum. The dynamics using a full Fourier transform of the spectral density also implies an anti-symmetrical form for $\mathcal{J}(\omega)$, that is:
\begin{equation}
\mathcal{J}(-\omega)=-\mathcal{J}(\omega)\,.
\end{equation}
Among the different functional forms that have been devised in the literature, we are hereafter referring to the so-called ohmic function, given by Eq.~(\ref{Eq:jw}), retaining but only two Lorentzians ($M=2$), enough to provide a well-structured form for this spectral density within the parameter range suggested in 
Ref\:\cite{tannor_meier_jw}. 
\begin{table}[ht]
	
	\tbl{Parameters of the spectral density}
	{\begin{tabular}{ccc} \toprule
			$\Delta_l$ $\,(\rm{a.u.})$    & $\omega_l$ (a.u.)   & $\Omega_l$ (a.u.) \\[0.5ex]
			\hline \noalign{\vskip 1.0ex}
			$1.0\times10^{-11} $    	     & $8.0\times10^{-4} $   	      & $1.4\times10^{-3} $    \\[0.5ex]
			$3.0\times10^{-12} $ 	     & $6.0\times10^{-3} $    	      & $4.0\times10^{-4} $    \\[0.5ex]
			\bottomrule
		\end{tabular}}
		\label{table:SD_values}
	\end{table}
The corresponding couplings and frequencies are gathered in Table \ref{table:SD_values}, whereas the resulting spectral density and the corresponding correlation function, at room temperature $300\,\,{\rm K} $ are illustrated in Fig. \ref{Fig:jw_ct}(a) and \ref{Fig:jw_ct}(b), respectively.

Two important arguments advocate for this specific choice of $\mathcal{J}(\omega)$:

i) A highly structured $\mathcal{J}(\omega)$ is in favor of enhanced memory effects in the bath response. In analogy with the Fano picture, such structures in $\mathcal{J}(\omega)$ could be attributed to some Feshbach resonances locally modifying the density of states. Actually they are supporting collective modes of the bath, and due to their possibly long enough lifetimes proportional to $\Omega_l^{-1}$ could temporarily trap the dynamics, leading to memory effects signatures.

ii) The central system transition frequency $\omega_0$ can be tuned in such an amount that it matches either one of the two maxima of the spectral density (case referred to as on-resonance) or, as an extreme situation, the minimum between the two peaks (case referred to as off-resonance). The off-resonance case is expected to produce the most important memory effects, as has already been pointed out in the literature \cite{tracedistanceSB} and as could be rationalized in terms of the back flow of information from the bath to the central system has to organized through the two neighbouring resonances. The specific two well separated peaks structure of the spectral density, offers a control flexibility by tuning the central transition frequency from on- to off-resonant cases. More specifically this could be achieved through an external laser field acting on the energy splitting of the central system via its transition dipole producing a controlled Stark shift.

\subsection{HEOM convergence.}

With the parameters that have finally been retained together with a bath temperature of $T=300K$, the overall dynamics is completely determined through the correlation function displayed in Fig. \ref{Fig:jw_ct}(b). It is interesting to note that $\mathcal{C}(t)$ shows about three damped oscillators of period $20 \,\,{\rm fs}$, with almost negligible values for times exceeding $100 \,\,{\rm fs}$, showing that memory effects can develop on ultra fast time scales. As has been explained in the previous theory section, this correlation function enters the memory kernel of the Nakajima-Zwanzig equation, which is further expanded in terms of successive HEOM levels ($L=1,2,\cdots)$. Our purpose is now to check the numerical convergence of the dynamical calculations as a function of increasing level of this hierarchy. For doing this we retain two observables on the reduced density matrix $\rho_S$, namely, its diagonal and off-diagonal terms $(\rho_S)_{11}$,$(\rho_S)_{12}$ and the volume of the accessible measuring the decoherence of the central system. The calculations are carried out fixing $\omega_0=4.0 \times 10^{-3} \,\,{ \rm a.u}$ which corresponds to an off-resonance case  close to the minimum between the two peaks of the spectral density displayed in Fig. \ref{Fig:jw_ct}(a).
 \begin{figure}[ht]
 	\centering
 	\includegraphics{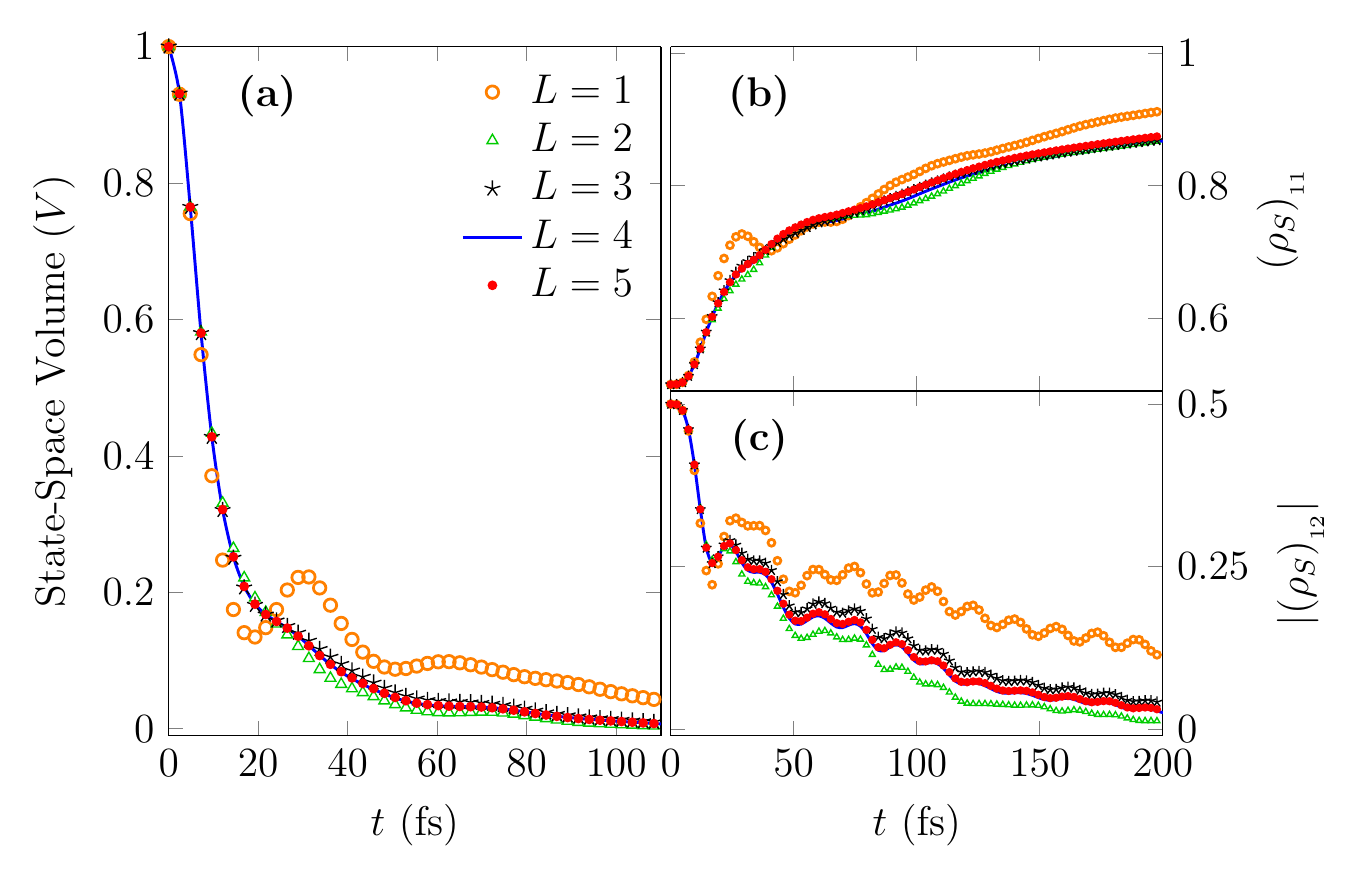}
 	\caption{(Color online)  Dynamics of the system calculated at different HEOM levels $L$. Panel (a)\:shows the state-Space volume $V(t)$, (b)\:depicts the diagonal element $(\rho_S)_{11}$ and (c)\: the off-diagonal element $\vert (\rho_S)_{12}\vert$ of the reduced density matrix $\rho_S$.}
 	\label{Fig:ConvergenceHEOM}
 \end{figure}

Fig. \ref{Fig:ConvergenceHEOM} gathers the results as a function of time, starting from a diabatic initial state where the whole population is in the central system ground state $(\rho_S)^d_{11}(0)=1$ in the diabatic representation, all other matrix elements of $\rho_S(0)$ being zero. The corresponding adiabatic picture, resulting from the diagonalization of $\hat{H}_S$, consists in taking all matrix elements $(\rho_S)_{ij}(0)=0.5$.   At HEOM $L=1$ level, often referred in the literature to be as converged for these rather low coupling regime (corresponding to the values of $\Delta_l$ listed in Table 1), we obtain a non-monotonic decay of the volume with, at least, two clearly identified bumps occurring at $t=30 \,\,{\rm fs}$ and $60 \,\,{\rm fs}$. Such bumps would be considered as convincing signatures of memory effects leading to non-Markovian behaviour. However, increasing HEOM level up to $L=4$ has as a consequence to partially wash out these bumps, fortunately still maintaining a non-monotonic decay especially observable for times longer than $20 \,\,{\rm fs}$. Finally, convergence is assumed to be satisfactory, as HEOM $L=5$ level gives almost the same results (see Fig. \ref{Fig:ConvergenceHEOM}) with reminiscences of the $L=1$ level bumps slightly shifted at times $t=30 \,\,{\rm fs}$ and $t=70 \,\,{\rm fs}$. Concerning the diagonal and off-diagonal matrix elements of $\rho_S(t)$ displayed in panels (b) and (c) of Fig. \ref{Fig:ConvergenceHEOM}, the converged calculations differ from the $L=1$ level one by erasing the shoulder at $30 \,\,{\rm fs}$. It is worthwhile noting that since $Tr[\rho_S(t)]=1$, we have accordingly checked that $(\rho_S)_{22} = 1 -(\rho_S)_{11}$, at any time $t$.  Off-diagonal elements asymptotically go to zero for the converged calculations. As a conclusion, for the forthcoming part of this work, we assume that $L=4$ is the HEOM level for which convergence is reached both for on- and off-resonance cases.

%!!!!!!!!!!!!!!!!!!!!!!!!!!!!!!!!!!!!!!!!!!!!!!!!!!!!!! %
\section{Results and Discussion}
\label{Sec:Results}
%!!!!!!!!!!!!!!!!!!!!!!!!!!!!!!!!!!!!!!!!!!!!!!!!!!!!!! %
The volume of the accessible states is decreasing as a function of time according to a time-dependent total decay rate $\Gamma(t)$, following an exponential law:
\begin{equation}
V(t)= det \vert F \vert=V(0) \exp [ -\frac{1}{2} \Gamma(t)t/\hbar]
\label{Eq:vol}
\end{equation}
as has previously been discussed \cite{relaxrates1}.
As opposite to a situation where the total decay rate is a constant, the above-mentioned law (Eq.~(\ref{Eq:vol})) induces a non-monotonic behaviour that could characterize non-Markovian dynamics. As has already been mentioned in the literature, a characteristic non-Markovian behaviour with a back-flow of information from the bath to the central system can be observed for temporarily negative values $\Gamma(t)$. 
Referring to converged calculations carried at HEOM $L=4$ level, we now examine two system-bath coupling situations, namely on- and off-resonant cases.

 \subsection{On-Resonant Case}
\begin{figure}[ht!]
 	\centering
 	 	\includegraphics{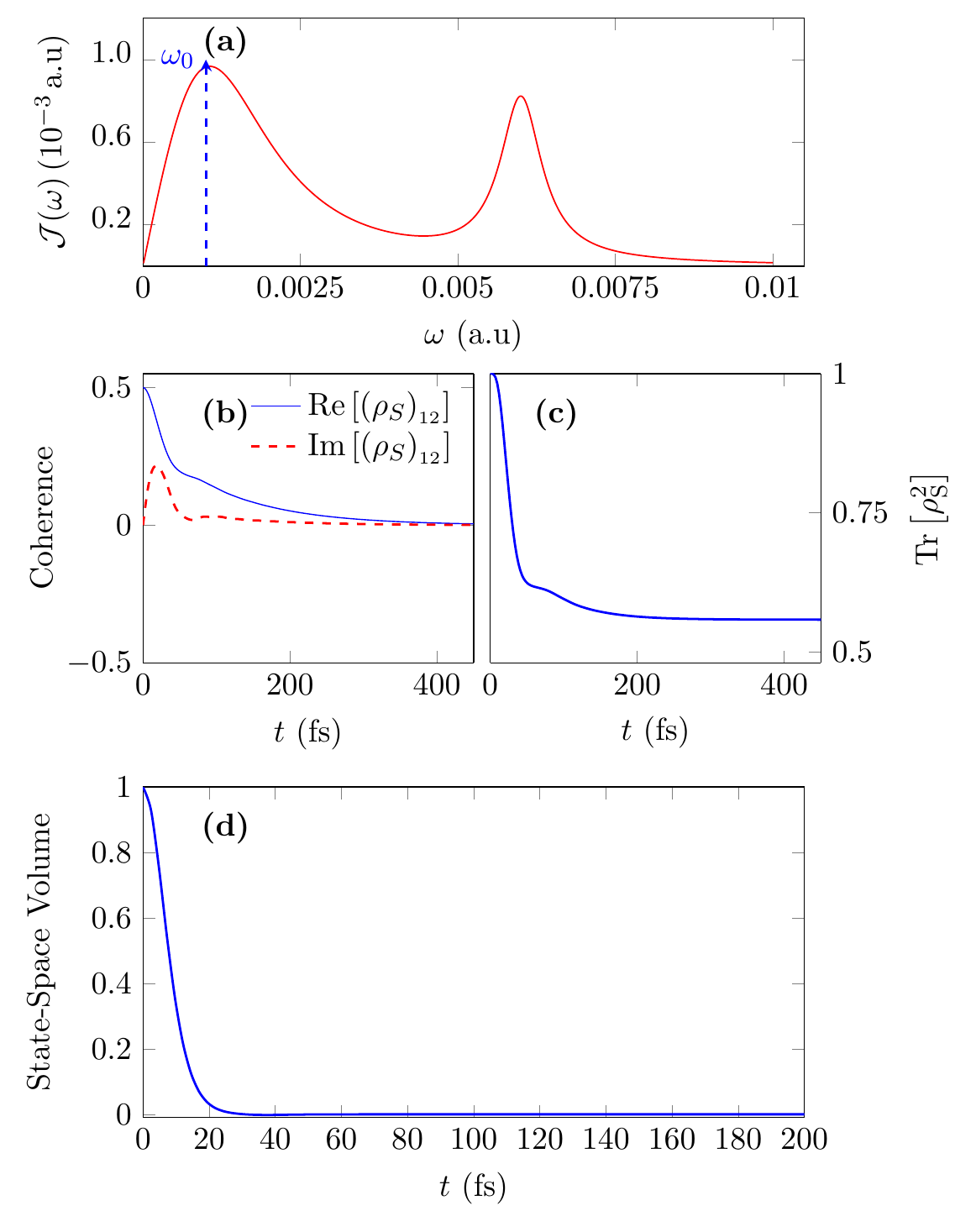}
 	\caption{(Color online)  Dynamics of the system for on-resonant case: Panel (a) shows the spectral density together with the position of the central system transition frequency $\omega_0$; Panels (b and c) display the off-diagonal element $(\rho_S)_{12}$ of the reduced density matrix and its trace Tr[$\rho_S^2$] and ; Panel (d) depicts the state-space volume $V(t)$.}
 	\label{Fig:on_Resonant}
 \end{figure}

 The central system internal transition frequency $\omega_0$ is tuned such as to match the first peak maximum of the spectral density, that is $\omega_0=1 \times 10^{-3} au$ as indicated in Fig.\ref{Fig:on_Resonant}(a).
 The results concerning the dynamical description of the reduced density matrix $\rho_S(t)$ are depicted as the trace Tr[$\rho_S^2$] in panel (c) and the real and imaginary parts of the off-diagonal elements $(\rho_S)_{12}$ in panel (b). The state-space volume is displayed in panel (d). We checked that the two curves resulting from Eq.~ (\ref{Eq:Vol_Access_state}) as the determinant of $F$, or from (Eq.~(\ref{Eq:vol})) as the decaying law with $\Gamma(t)$, are perfectly superimposed. The initial values are 1 for Tr[$\rho_S^2(0)$] and the volume $V(0)$, and 0.5 for the off-diagonal elements $(\rho_S)_{12}(0)$. Basically two decoherence time scales are observed. The first, of about $200 \,\,{\rm fs}$, concerns the decay of Tr[$\rho_S^2$] and $(\rho_S)_{12}$ both showing rather monotonic behaviour, with a shoulder at about $100 \,\,{\rm fs}$ and a small amplitude oscillation for the imaginary part of $(\rho_S)_{12}$ at about $20 \,\,{\rm fs}$. Tr[$\rho_S^2$] finally reaches its asymptotic value of 0.5. The second time scale, much shorter, characterizes the dynamics of the volume of the accessible states which is decaying almost monotonously within about $20 \,\,{\rm fs}$ (panel d). This is not only a signature of a fast dynamical evolution (typically a vibrational period), but also of a presumably Markovian, or close to Markovian behaviour. 
 
 For this resonant case our conclusion is that the volume of accessible states space is not by itself a probe sensitive enough to detect a clear non-Markovian behaviour. 
 
  \subsection{Off-Resonant Case}
\begin{figure}[ht!]
  	\centering
  \includegraphics{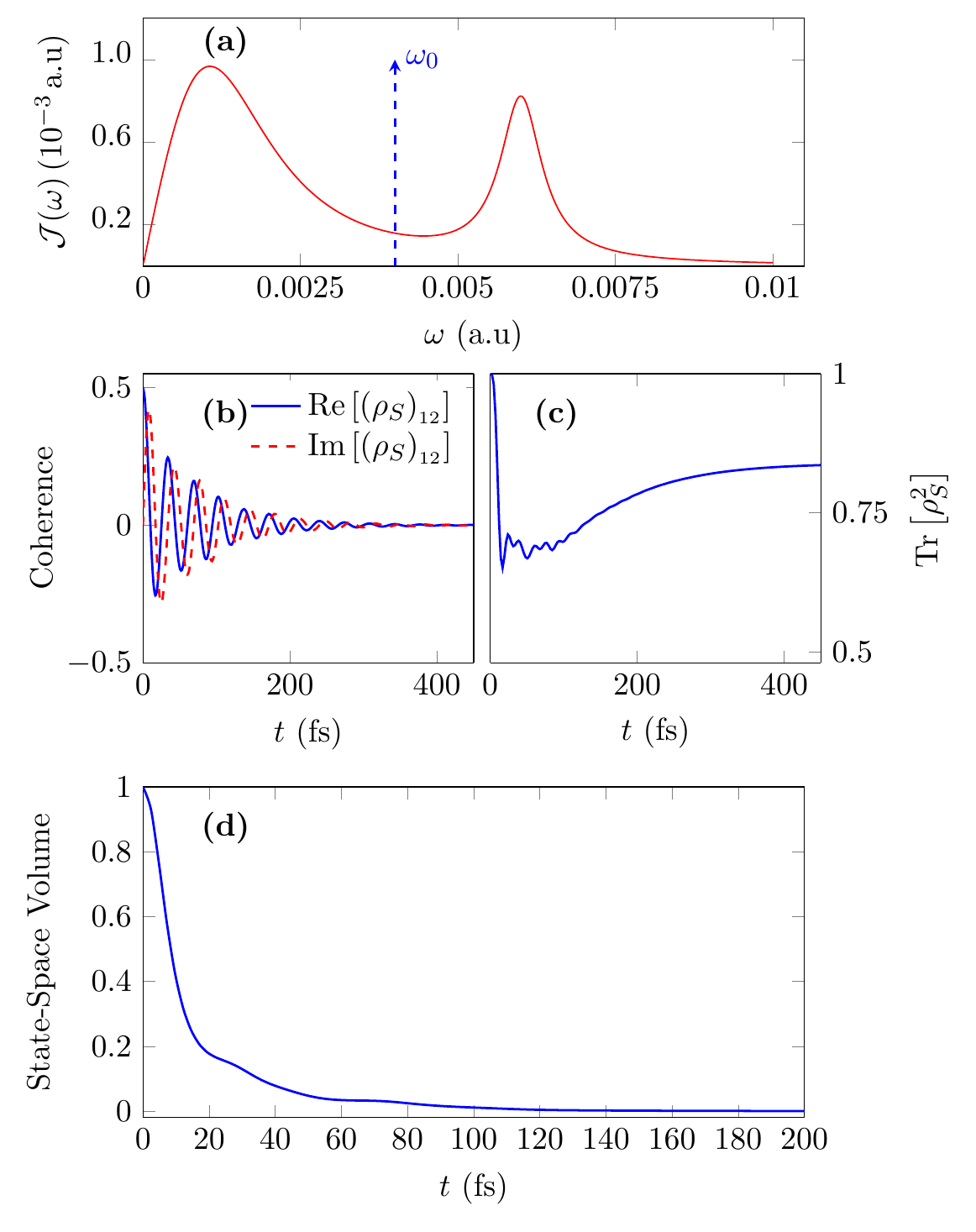}
  	\caption{(Color online) Same as in figure 3 but for the off-resonant case.}
  	\label{Fig:Off_Resonant}
  \end{figure}

  The central system internal transition frequency $\omega_0$ in now tuned off-resonance around the minimum of the spectral density $\omega_0=4 \times 10^{-3}\,\,{\rm a.u}$ (see Fig. \ref{Fig:Off_Resonant}(a)). As resulting from the discussion of the model, this situation is expected to support better marked non-Markovianity. As in the on-resonant case, two decoherence time scales are also observed here. The long one, characterizing Tr[$\rho_S^2$] and $(\rho_S)_{12}$, is still about $200 \,\,{\rm fs}$. The difference being that both dynamics are not monotonic and the asymptotic value of Tr[$\rho_S^2$] is now much larger (more than 0.8) showing a better preservation of the central system, and even more interestingly, an increasing behaviour starting from about $100 \,\,{\rm fs}$ (Fig. \ref{Fig:Off_Resonant}(c) and (b) and supporting a back flow of information from the bath. As for the observable taken as the signature of non-Markovianity, the volume of accessible states $V(t)$ is displayed in Panel (d), with the second characteristic shorter time-scale. Contrary to the on-resonant case, a clear non-monotonic behaviour is obtained with a small bump at $t=30 \,\,{\rm fs}$, the overall complete decay occurring at $60\,\,{\rm fs}$. In other words, the characteristic decoherence time has been slowed down from $20 \,\,{\rm fs}$ to $60 \,\,{\rm fs}$.
  
  For this off-resonant case, our conclusion is that, as expected the non-Markovianity is better marked on the volume total decay rate.
  
  Finally, the control strategy, as suggested previously, is to adequately tune the central system internal transition frequency from on- to off-resonant cases, in order to increase the non-Markovian behaviour. This can be achieved using a strong external laser field inducing a Stark shift between the levels of the central system.
%!!!!!!!!!!!!!!!!!!!!!!!!!!!!!!!!!!!!!!!!!!!!!!!!!!!!!! %
\section{Conclusion and Outlook}\label{Sec:Conclusion}
Different laser control scenarios could be looked for when dealing with open quantum systems with the potentiality to be applied to a broad range of processes from biology (proton transfer in large proteins), to chemistry (charge transfers in donor-acceptor systems) and physics (molecular alignment-orientation, quantum information). Such scenarios can roughly be classified in two groups according to the situation where the control field is acting on the central two-level system alone (by modifying its internal characteristic transition frequency) or on the whole system including the bath (by modifying appropriate collective modes responsible for the above mentioned physical processes). In the present work we are only referring to the simplest control scenario of the first class with a control field just producing a Stark shift  on the dressed levels of the central system with a consequence on its coupling scheme with the bath. To the best of our knowledge, this is one of the very first attempts toward control in such complex systems. 

The model in consideration is a standard Spin-Boson Hamiltonian with a double-peaked spectral density function. The calculations are carried by solving a Nakajima-Zwanzig type master equation with a converged HEOM level for the memory kernel. More precisely, for the chosen parameters of the mode, the central system typically supporting a molecular vibrational mode coupled to a bath at an equilibrium room temperature, fully converged calculations need not less than level 4 for the HEOM hierarchy. The external field intensity is continuously varied to tune the transition frequency of the central system from on- to off-resonant situations, that are reached when this frequency matches respectively the maximum or minimum of the spectral density. The observable measuring the overall decoherence of the two-level system towards the bath taken into account is the time-dependent decay of the accessible states mapping the central system reduced density matrix. A non-monotonic behaviour is considered as a possible signature of non-Markovianity. We numerically show and rationalize that the on-resonant case corresponds to basically memory-less, fast dynamics, whereas the off-resonant configuration may bring into play better marked memory effects and longer decoherence times.

As a word of conclusion, a simple control can be exerted merely by an appropriate tuning of the transition frequency to enhance non-Markovian bath response. The resulting longer decoherence times may advantageously be exploited to enhance the efficiency and long period robustness of molecular alignment-orientation or qubits information preservation. Moreover, very challenging perspectives are now opening with more sophisticated control strategies aiming at optimizing and coherently interfering bath collective modes through appropriate combinations of the eigen-channels of the decoherence dynamics \cite{PhysRevA.89.062310, Reich2015, Glaser2015, chenel, JChemPhys.Michele, NJP.Schirmer}. Ultimately, fast and efficient charge or proton transfers in long protein chains could be considered. We are actively pursuing our investigations along these lines.

\section*{Acknowledgments}
We acknowledge Prof. Christiane Koch, Prof. Christoph Meier, Prof. Arne Keller and Prof. Eric Charron for fruitful discussions. O.A. thanks the organizers of the \emph{Molecules and Laser Fields Symposium} in honour of Andre Bandrauk, in Orford (QC), Canada, May 4-7, 2016, where part of the control scheme has been discussed. We acknowledge support from the ANR-DFG, under Grant No. \mbox{ANR-15-CE30-0023-01.} This work has been performed with the support of the Technische Universit\"at M\"unchen Institute for Advanced Study, funded by the German Excellence Initiative and the European Union Seventh Framework Programme under Grant Agreement No. 291763.


\begin{thebibliography}{56}
	
	\bibitem{breuerbook}
	H.P. Breuer and F.~Petruccione
	\newblock \emph{The Theory of Open Quantum Systems}
	\newblock Oxford University Press (2002)
	
	\bibitem{weissbook}
	U.~Weiss
	\newblock \emph{Quantum Dissipative Systems}
	\newblock World Scientific (1999)
	
	\bibitem{RMPVega}
	I.~de Vega and D. Alonso
	\newblock \emph{In Press: Rev. of. Mod. Phys} (2016)
	\newblock arXiv:1511.06994.
	
	\bibitem{Chin2013}
	A.~W. Chin, J.~Prior, R.~Rosenbach, F.~Caycedo-Soler, S.~F. Huelga and M.~B. Plenio
	\newblock \emph{Nat. Phys.}, \textbf{9} 113-118 (2013)
	
	\bibitem{Gelinas2014}
	S.G{\'e}linas, A.Rao, A. Kumar, S.~L. Smith, A.~W. Chin, J. Clark, T.~S. van~der Poll, G.~C. Bazan and R.~H. Friend
	\newblock \emph{Science}, \textbf{343} 512-516 (2014)
	
	\bibitem{Scholes2011}
	G.~D. Scholes, G.~R. Fleming, A. Olaya-Castro and R. van Grondelle
	\newblock \emph{Nat. Chem.}, \textbf{3} 763-774 (2011)
	
	\bibitem[Meyer et~al.(2009)Meyer, Gatti and Worth]{GattiMCTDH}
	H.-D. Meyer, F. Gatti and G.~A. Worth
	\newblock \emph{Multidimensional Quantum Dynamics: MCTDH Theory and	Applications}
	\newblock Wiley (2009)
	
	\bibitem[Gaspard and Nagaoka(1999)]{stochastic}
	P.~Gaspard and M.~Nagaoka
	\newblock \emph{J. Chem. Phys.}, \textbf{111} 5676-5690 (1999)
	
	\bibitem{Bandrauk}
	S. Miret-Art\'es, O. Atabek and A.~D. Bandrauk
	\newblock \emph{Phys. Rev. A}, \textbf{45} 8056-8063 (1992)
	
	\bibitem{Lindblad1}
	V. Gorini, A. Kossakowski and E.~C.~G. Sudarshan
	\newblock \emph{J. Math. Phys.}, \textbf{17} 821 (1976)
	
	\bibitem{Lindblad2}
	G.~Lindblad
	\newblock \emph{Com. Math. Phys.}, \textbf{48} 119 (1976)
	
	\bibitem{KochReview}
	C.~P. Koch
	\newblock \emph{J. Phys. Cond. Mat.}, \textbf{28} 213001 (2016)
	
	\bibitem{collectivemode}
	I. Burghardt, R. Martinazzo and K.~H. Hughes
	\newblock \emph{J. Chem. Phys.}, \textbf{137} (2012)
	
	\bibitem{tracedistance}
	H.-P. Breuer, E.-M. Laine and J. Piilo
	\newblock \emph{Phys. Rev. Lett.}, \textbf{103} 210401 (2009)
	
	\bibitem{tracedistance2}
	E.-M. Laine, J. Piilo and H.-P. Breuer
	\newblock \emph{Phys. Rev. A}, \textbf{81} 062115 (2010)
	
	\bibitem{PhysRevA.88.020102}
	S. Lorenzo, F. Plastina and M. Paternostro
	\newblock \emph{Phys. Rev. A}, \textbf{88} 020102 (2013)
	
	\bibitem{relaxrates1}
	M. J.~W. Hall, J.~D. Cresser, L.~Li and E. Andersson
	\newblock \emph{Phys. Rev. A}, \textbf{89} 042120 (2014)
	
	\bibitem{heom1}
	Y. Tanimura and R. Kubo
	\newblock \emph{J. Phys. Soc. Japan}, \textbf{58}	101-114 (1989)
	
	\bibitem{heom2}
	A. Ishizaki and Y. Tanimura
	\newblock \emph{J. Phys. Soc. Japan}, \textbf{74} 3131-3134 (2005)
	
	\bibitem{jcp126.2538754}
	M. Schr\"oder, M. Schreiber and U. Kleinekath\"ofer
	\newblock \emph{J. Chem. Phys.}, \textbf{126} 114102 (2007)
	
	\bibitem{jcp130.3155372}
	A. Ishizaki and G.~R. Fleming
	\newblock \emph{J. Chem. Phys.}, \textbf{130} 234111 (2009)
	
	\bibitem{PhysRevE.86.011915}
	A.~Shabani, M.~Mohseni, H.~Rabitz and S.~Lloyd
	\newblock \emph{Phys. Rev. E}, \textbf{86} 011915 (2012)
	
	\bibitem{RMPbreuer}
	H.-P. Breuer, E.-M. Laine, J. Piilo and B. Vacchini
	\newblock \emph{Rev. Mod. Phys.}, \textbf{88} 021002 (2016)
	
	\bibitem{reviewrivas}
	\'A. Rivas, S.~F. Huelga and M.~B. Plenio
	\newblock \emph{Rep. Prog. Phys.}, \textbf{77} 094001 (2014)
	
	\bibitem{PhysRevLett.105.050403}
	\'A. Rivas, S.~F. Huelga and M.~B. Plenio
	\newblock \emph{Phys. Rev. Lett.}, \textbf{105} 050403 (2010)
	
	\bibitem{PhysRevA.84.052118}
	R. Vasile, S. Maniscalco, M. G.~A. Paris, H.-P. Breuer and J. Piilo
	\newblock \emph{Phys. Rev. A}, \textbf{84} 052118 (2011)
	
	\bibitem{PhysRevA.87.042103}
	J. Liu, X.-M. Lu and X. Wang
	\newblock \emph{Phys. Rev. A}, \textbf{87} 042103 (2013)
	
	\bibitem{PhysRevA.82.042103}
	X.-M. Lu, X. Wang and C.~P. Sun
	\newblock \emph{Phys. Rev. A}, \textbf{82} 042103 (2010)
	
	\bibitem{PhysRevLett.112.120404}
	D. Chru\ifmmode \acute{s}\else \'{s}\fi{}ci\ifmmode~\acute{n}\else\'{n}\fi{}ski and S. Maniscalco
	\newblock \emph{Phys. Rev. Lett.}, \textbf{112} 120404 (2014)
	
	\bibitem{mesure_expnature}
	J.~T. Barreiro, M. Muller, P. Schindler, D. Nigg, T. Monz,	M. Chwalla, M. Hennrich, C.~F. Roos, P. Zoller and R. Blatt
	\newblock \emph{Nature}, \textbf{470} 486-491 (2011)
	
	\bibitem{mesure_expnaturephys}
	B.-H. Liu, L.~Li, Y.-F. Huang, C.-F. Li, G.-C. Guo, E.-M. Laine, H.-P. Breuer and J. Piilo
	\newblock \emph{Nat. Phys.}, \textbf{7} 931-934 (2011)
	
	\bibitem{tracedistanceSB}
	G. Clos and H.-P. Breuer
	\newblock \emph{Phys. Rev. A.}, \textbf{86} 012115 (2012)
	
	\bibitem{nonmarkoviansaalfrank}
	U. Lorenz  and P. Saalfrank
	\newblock \emph{Eur. Phys. J. D.}, \textbf{69} 46 (2015)
	
	\bibitem{JChemPhys.134.101103}
	P. Rebentrost and A. Aspuru-Guzik
	\newblock \emph{J. Chem. Phys.}, \textbf{134} 101103 (2011)
	
	\bibitem{relaxrates2}
	E. Andersson, J.~D. Cresser and M.~J.~W. Hall
	\newblock \emph{J. Mod. Opt.}, \textbf{54} 1695-1716 (2007)
	
	\bibitem{localnonlocal}
	D. Chru\ifmmode \acute{s}\else \'{s}\fi{}ci\ifmmode~\acute{n}\else\'{n}\fi{}ski and A. Kossakowski
	\newblock \emph{Phys. Rev. Lett.}, \textbf{104} 070406 (2010)
	
	\bibitem{Nakajima01121958}
	S. Nakajima
	\newblock \emph{Prog.Theo. Phys.}, \textbf{20} 948-959 (1958)
	
	\bibitem{ZWANZIG19641109}
	R. Zwanzig
	\newblock \emph{Physica}, \textbf{30} 1109-1123 (1964)
	
	\bibitem{Shibata1977}
	F. Shibata, Y. Takahashi and N. Hashitsume
	\newblock \emph{J. Stat. Phys.}, \textbf{17} 171-187 (1977)
	
	\bibitem{Hashitsumae1977}
	N. Hashitsumae, F. Shibata and M. Shingu
	\newblock \emph{J. Stat. Phys.}, \textbf{17} 155-169 (1977)
	
	\bibitem{FEYNMAN1963118}
	R.P Feynman and F.L Vernon
	\newblock \emph{Annals of Physics}, \textbf{24} 118-173 (1963)
	
	\bibitem{Makri95}
	N. Makri
	\newblock \emph{J. Math. Phys.}, \textbf{36} 2430-2457 (1995)
	
	\bibitem{Makri99}
	N. Makri
	\newblock \emph{J. Chem. Phys.}, \textbf{111} 6164-6167 (1999)
	
	\bibitem{PhysRevA.41.6676}
	Y. Tanimura
	\newblock \emph{Phys. Rev. A}, \textbf{41} 6676-6687 (1990)
	
	\bibitem{JChemPhys.122.041103}
	R.-X. Xu, P. Cui, X.-Q. Li, Y. Mo and Y. Yan
	\newblock \emph{J. Chem. Phys.}, \textbf{122}	041103 (2005)
	
	\bibitem{jcp140.4870035}
	Hao Liu, Lili Zhu, Shuming Bai and Qiang Shi
	\newblock \emph{J. Chem. Phys.}, \textbf{140} 134106 (2014)
	
	\bibitem{tannor_meier_jw}
	C. Meier and D.~J. Tannor
	\newblock \emph{J. Chem. Phys.}, \textbf{111} 3365-3376 (1999)
	
	\bibitem{Pomyalov201098}
	A.~Pomyalov, C.~Meier and D.J. Tannor
	\newblock \emph{Chemical Physics}, \textbf{370} 98-108	(2010)
	
	\bibitem{NIETO199554}
	A. Nieto
	\newblock \emph{Comp. Phys. Comm.}, \textbf{92} 54-64 (1995)
	
	\bibitem{JChemTeoComp.8.ct3003833}
	J. Str\"umpfer and K. Schulten
	\newblock \emph{J. Chem. Theo. Comp.}, \textbf{8} 2808-2816 (2012)
	
	\bibitem{PhysRevA.89.062310}
	J.-S. Tai, K.-T. Lin and H.-S. Goan
	\newblock \emph{Phys. Rev. A}, \textbf{89} 062310 (2014)
	
	\bibitem{Reich2015}
	D.~M. Reich, N. Katz and C.~P. Koch
	\newblock \emph{Sci. Rep.}, \textbf{5} 12430 (2015)
	
	\bibitem{Glaser2015}
	S.~J. Glaser, U. Boscain, T. Calarco, C.~P. Koch, W. K{\"o}ckenberger, R. Kosloff, I. Kuprov, B. Luy, S. Schirmer,	T. Schulte-Herbr{\"u}ggen, D. Sugny and F.~K. Wilhelm
	\newblock \emph{Eur. Phys. J. D.}, \textbf{69} 279 (2015)
	
	\bibitem{chenel}
	A.~Chenel, G.~Dive, C.~Meier and M.~Desouter-Lecomte
	\newblock \emph{J. Phys. Chem. A.}, \textbf{116} 11273-11282 (2012)
	
	\bibitem{JChemPhys.Michele}
	A.~Chenel, C.~Meier, G.~Dive and M.~Desouter-Lecomte
	\newblock \emph{J. Chem. Phys.}, \textbf{142} 024307 (2015)
	
	\bibitem{NJP.Schirmer}
	F. F. Floether, P. de~Fouquieres and S. G. Schirmer
	\newblock \emph{New J. Phys.}, \textbf{14} 073023 (2012)
	
\end{thebibliography}
\end{document}